\documentclass[twocolumn]{aastex701}
\usepackage{hyperref}
\usepackage{wrapfig}
\usepackage{caption}
\usepackage{subcaption}
\usepackage{graphicx}
\usepackage{ragged2e}
\usepackage{wasysym}
\usepackage{booktabs}
\usepackage{enumitem}

\newcommand{\NU}{
Department of Physics and Astronomy, Northwestern University, Evanston, IL, USA}
\newcommand{\CIERA}{
Center for Interdisciplinary Exploration and Research in Astrophysics, Northwestern University, Evanston, IL, USA}
\newcommand{\SkAI}{
NSF-Simons AI Institute for the Sky (SkAI), 172 E. Chestnut St., Chicago, IL 60611, USA}

\def\pfrac#1#2{\left( \frac{#1}{#2} \right)}
\def\volrate{\mathcal{R}}
\def\asymuncert#1#2#3{#1^{+#2}_{-#3}}

\begin{document}

\title{~~\\
Searching for Neutron Star Mergers in the Absence of Gravitational Waves \\ with Optical Afterglow Emission
~~\\}

\author[0009-0000-5561-9116]{Haille M. L. Perkins}
\email{haillep2@illinois.edu}
\affiliation{Department of Astronomy, University of Illinois Urbana-Champaign, Urbana, IL 61801}
\affiliation{Center for Astrophysical Surveys, National Center for Supercomputing Applications, Urbana, IL 61801}
\affiliation{Illinois Center for Advanced Studies of the Universe, University of Illinois Urbana-Champaign, Urbana, IL 61801}

\author[0000-0001-6022-0484]{Gautham Narayan}
\email{gsn@illinos.edu}
\affiliation{Department of Astronomy, University of Illinois Urbana-Champaign, Urbana, IL 61801}
\affiliation{Center for Astrophysical Surveys, National Center for Supercomputing Applications, Urbana, IL 61801}
\affiliation{Illinois Center for Advanced Studies of the Universe, University of Illinois Urbana-Champaign, Urbana, IL 61801}
\affiliation{\SkAI}

\author[0000-0002-4188-7141]{Brian D. Fields}
\email{bdfields@illinos.edu}
\affiliation{Department of Astronomy, University of Illinois Urbana-Champaign, Urbana, IL 61801}
\affiliation{Illinois Center for Advanced Studies of the Universe, University of Illinois Urbana-Champaign, Urbana, IL 61801}
\affiliation{Department of Physics, University of Illinois Urbana-Champaign, Urbana, IL 61801}

\author[0009-0009-1590-2318]{Ved G. Shah}
\email{vedshah2029@u.northwestern.edu}
\affiliation{Department of Astronomy, University of Illinois Urbana-Champaign, Urbana, IL 61801}
\affiliation{\NU}
\affiliation{\CIERA}
\affiliation{\SkAI}

\author[0000-0001-9915-8147]{Genevieve~Schroeder}
\email{gms279@cornell.edu}
\affiliation{Department of Astronomy, Cornell University, Ithaca, NY 14853, USA}

\correspondingauthor{Haille Perkins}
\email{haillep2@illinois.edu}

\begin{abstract}
With the forth observing run of the LIGO–Virgo–KAGRA gravitational-wave network, which enabled the discovery of the kilonova (KN) counterpart to GW170817, ending with no new confirmed neutron star mergers, the intrinsic rate of these events must be even lower than previously estimated. As a result, building a sample of KNe will remain challenging even with continued GW observations, motivating complementary discovery strategies that do not rely on gravitational-wave triggers.  In this work, we consider how leveraging bright short gamma-ray burst afterglows can aid in the discovery on KNe with the Rubin Observatory's upcoming Legacy Survey of Space and Time (LSST),  whose unprecedented depth will make such detections feasible.  We find that nearly on-axis ($\theta_{\rm view} \leq 30^{\circ}$) afterglows can enhance KN detection rates in the LSST $g$-band from $\asymuncert{29}{51}{21} \ \rm yr^{-1}$ to $\asymuncert{91}{160}{65} \ \rm yr^{-1}$.  We further show how the colors of the observed events can be used to distinguish between neutron star merger counterparts with and without KN emission.  This study demonstrates how critical multi-wavelength and multi-survey observations are for these rare events, especially without context from gravitational waves. Fortunately, detectable events will likely be discovered near peak with LSST, allowing for rapid follow-up and confirmation. We discuss key uncertainties in our study, particularly volume rate of merger events, and the degeneracy between the empirically determined explosion energy and ambient medium density.
\end{abstract}


\section{Introduction} \label{sec:intro}

Binary neutron star merger (NSM) event GW170817 was first detected via its gravitational wave (GW) signal \citep{Abbott_2017_gw} and was also found coincident with a gamma-ray burst, GRB170817A, making it the first event have been detected in gravitational and electromagnetic radiation \citep{Abbott_2017_gwgrb}.  This joint detection prompted a large, multi-facility, multi-wavelength observational campaign in search for a kilonova (KN) counterpart -- the thermal transient arising from the decay of the heavy elements synthesized in the blast \cite[e.g.][]{metzger_electromagnetic_2010}, \cite[e.g.][]{abbott_2017_mm}.  The search was successful; in just $11$ hrs post-merger, SSS17a/AT2017gfo was identified, making it the only confirmed NSM with an observed KN. Moreover, faster processing of the gravitational wave skymap would have likely yielded an even earlier discovery.

Extensive follow-up and modeling of the KN suggests the event was consistent with the production of {\it r}-process, or rapid neutron capture process, material \citep{villar_combined_2017}, which was predicted decades prior \cite[e.g.][]{lattimer_black-hole-neutron-star_1974}. NSMs are believed to be one of the main site of $r$-process production \citep{chen_neutron_2024}, as they can construct the neutron-rich environment required for nuclei to capture neutrons at a timescale shorter than that of $\beta$-decay \citep{burbidge_synthesis_1957, Cameron_1957, Cowan_2021}.  While other sites, including recently confirmed magnetar giant flares \citep{Patel2025_magFlares} along with theorized sites such as special classes of core-collapse supernovae (e.g. collapsars, \citealt{Agarwal2025_collapsars}; magnetorotational supernovae \citealt{Reichert2023_mrsne}; common envelope jet supernovae, \citealt{Grichener2025_CESNe}), may contribute to the $r$-process budget of the universe, determining what fraction is produced by NSM is of interest.  NSMs also have implications in multi-messenger astronomy \citep[MMA; ][]{Pian_2020}, the nature of neutron stars \citep{Abbott_2019}, heavy element nucleosynthesis \citep[e.g.][]{Arnett1996, Lattimer1974}, galactic chemical evolution \citep[e.g.][]{Koo_2020feedback, Kobayashi_2020gce}, cosmology \citep{Abbott_2017_h0}, and more. There are several candidate KN from GRB follow-up, such as GRBs 211211A \citep{Rastinejad_2022_grb211211, troja_nearby_2022} and 230307A \citep{Bulla_2023_grb23, Gillanders_2023_grb23, Levan_2023_grb23}; however, since GW170817, there has not been another bonafide NSM discovery.  With the lack of an observed population, further observations will not only provide insights into the intrinsic variation of such events but also support a wide breadth of complementary science cases. 
 
While the gravitational wave signal was crucial for enabling rapid follow-up of AT2017gfo, several studies have investigated the prospects of joint GW and electromagnetic detections throughout LVK observing run 04. Selecting the most optimistic estimates from \citet{Mochkovitch_2021_o4rates, Frostig_2022_winter, shah_predictions_2024} ($7.9 \rm \ yr^{-1}$, $\asymuncert{6}{3}{4}$, $\asymuncert{2}{3}{2}$, respectively) highlights that, even under favorable assumptions, expected detections remain small.  With 04 now concluded, without any further NSMs, these estimates were truly optimistic and the intrinsic rate of NSM is much lower than prior estimates.  Although Virgo and KAGRA participated in O4, their sensitivities were significantly lower than that of LIGO, limiting their contribution primarily to constraining non-detections. As a result, even with a global GW detector network in operation, the intrinsic rarity of NSMs fundamentally limits the number of observable kilonovae.  Looking ahead, a planned six-month extension of O4 (O4d)\footnote{As of November 18, 2025, plans for O4d remain provisional. \url{https://observing.docs.ligo.org/plan/}  Unless otherwise specified, information regarding LVK sensitivities and observing runs is obtained from this site.} in mid-to-late 2026 is expected to overlap with the start of operations of the Rubin Observatory's optical Legacy Survey of Space and Time \citep[LSST;][]{Ivezic2019_lsst}. While this overlap will improve joint discovery prospects relative to earlier phases of O4, the underlying challenge remains the intrinsically low event rate of NSMs. Consequently, building a statistically meaningful sample of KNe will require complementary discovery channels beyond GW-triggered searches alone.

In the absence of frequent GW triggers, wide-field astronomical surveys provide a promising avenue for identifying KN candidates. Although KN emission is relatively isotropic, these transients are intrinsically faint compared to typical transients, such as supernovae. This, coupled with their intrinsic rarity (with most recent rate estimate of $\asymuncert{56}{99}{40} \rm \ Gpc^{-3} yr^{-1}$ \citep{Akyuz2025rate}), further increases the difficulty of discovering such events. Given these limitations, we investigate a means of finding candidate events without GW triggers. To do this, we considered leveraging another electromagnetic counterpart to NSMs: short $\gamma$-ray burst afterglows (sGRB AG).  The sGRB, an energetic burst of gamma rays from the highly relativistic jet launched from the merger, shocks the surrounding interstellar medium producing an AG of synchrotron emission \citep[e.g.][]{Sari_1998_afterglow} visible across the electromagnetic spectrum.  This work explores how coincident emission from KNe and AGs may impact the detection of NSM in the upcoming wide-field optical Legacy Survey of Space and Time \citep{Ivezic2019_lsst}. 

Given the intrinsic rarity of NSM and faintness of KNe, discovery without GW context will require a survey with rapid and deep observations across large regions of the sky.  \cite{Andreoni2021_ztfkne} made initial attempts of KNe searches with the Zwickey Transient Facility (ZTF), which is sensitive to these events out to $200 \ \rm Mpc$.  They developed the \texttt{ZTFReST} infrastructure to search for fast evolving candidate transients in both archival and real time data leading to the discovery of 3 confirmed new AGs but no KNe.  Other searches include the optical KNTraP \citep{VanBemmel_2025_kntrap} and infrared WINTER \citep{Frostig_2022_winter}, but also without any KNe discoveries.  Although, we can look forward to the Rubin Observatory's LSST, which is set to begin early 2026\footnote{\url{https://rtn-011.lsst.io/}}.  LSST will cover nearly all of the southern sky every few nights in 6 bands ($ugrizY$), compared to 2 ($gr$) of the public ZTF survey.  Additionally, LSST will be much deeper, capable of detecting GW170817-like events out to $\sim\!600 \ \rm Mpc$ expanding the search volume by a factor of 27 per solid angle relative to ZTF. This extra volume is critical when searching for such rare events.  

While KN emission can be faint, the AG is another source of optical emission.  In the case of GW170817, the AG was seen across the electromagnetic spectrum \citep[e.g.][]{troja_x-ray_2017, makhathini_panchromatic_2021}; however, the event was viewed $\sim20 \ \rm deg$ from the polar axis \citep{Finstad_2018thetav, Abbott_2019, ryan_gamma-ray_2020, makhathini_panchromatic_2021}, and as a result there was a delay in the peak AG emission until $\sim 1$ yr after the merger. If the event had been viewed on-axis, the AG would have been brighter than the KN and peaked on a similar timescale \citep{Salafia_2019A17onaxis, zhu_kilonova_2022}.  There have even been a few cases where candidate KNe were found alongside AGs of long GRB detections \citep{Rastinejad_2022_grb211211, Levan_2023_grb23, Gillanders_2023_grb23}, demonstrating the utility of AG observations.   

The paper is structured as follows:  Section \ref{sec:synthetic_sed} describes the physical parameters necessary to generate  the synthetic SEDs for both the KN and the AG. In Section \ref{sec:simulate}, we describe the simulation of a sample of KN and AGs and assess the typical appearance of the coincident emission in Section \ref{sec:combinedLightcurves}. Section \ref{sec:discovery} highlights prospects of discovering these events in the upcoming Legacy Survey of Space and Time and the expected rate of observing them. 
Finally, an overall summary and future considerations of AG enhancement are found in Section \ref{sec:summary}.

\section{Including Afterglow Emission} \label{sec:synthetic_sed}

Here, we describe the generation of spectral energy distributions (SEDs) for both KN and AG using existing models \citep{bulla_possis_2019, ryan_gamma-ray_2020} which are overlaid to investigate the coincident emission. We will then use the simulated SEDs to construct synthetic light curves with \texttt{sncosmo} \citep{sncosmo}.

\subsection{Kilonova SEDs}

The observed KN is influenced by many physical factors, including the abundance, composition, and distribution of ejected material \citep[e.g.][]{li_transient_1998, metzger_electromagnetic_2010}.  Each KN model accounts for these differently \citep[e.g.][]{Kasen_2015, villar_combined_2017, Hotokezaka_2020, metzger_kilonovae_2020}. In this work, we generate our KN SEDs following the method developed in \cite{shah_predictions_2024}, which interpolates over an existing finite grid of KN SEDs \citep{dietrich_multi-messenger_2020} using physically motivated scaling laws. The grids were simulated using \texttt{POSSIS} \citep{bulla_possis_2019, Bulla_2023_possis}, a time-dependent, 3-dimensional Monte Carlo radiative transfer code \citep{dietrich_multi-messenger_2020}, which assumes the physical description of the event as detailed below.

The ejecta from a NSM comes in two dominant forms \citep[e.g.][]{Fernandez_2016_nsms, metzger_kilonovae_2020}. The first being the material lost due to the tidal interactions, known as the dynamical ejecta. Some material does not escape immediately and forms a disk around the compact remnant and the outflows from this disk are known as the disk wind \citep{Metzger_2008_disks}.  The \cite{dietrich_multi-messenger_2020} grids account for this in two parameters, the mass of the dynamical ejecta, $m^{\rm dyn}_{\rm ej}$, and disk-wind ejecta, $m^{\rm wind}_{\rm ej}$.  The model further assumes that due to potential neutrino irradiation from the compact object, the dynamical ejecta is divided into two regions. The neutrino irradiation tends to constrain the heavy $r$-process (lanthanide-rich) material in a region close to the plane of the merger with an opening angle of $\Phi$. This describes the region in velocity-space for which the ejecta is lanthanide-rich. As a result, there is a viewing angle dependence as regions with higher abundances of $r$-process material have greater opacity, thus producing a redder transient \citep[e.g.][]{Barnes2013_kne, Metzger_2014_redblue, Perego_2017_anisotropic}. This viewing angle is encoded in the final model parameter: the cosine of the viewing angle, $\cos \theta_{\rm view}$.  The four model parameters, along with the distributions we sample them from, are described in Table \ref{table:KN-parameters}.

As mentioned, there are a finite number of grids simulated that cover discrete selections within the 4-D parameter space \citep{dietrich_multi-messenger_2020}, so to evaluate at arbitrary combinations of parameters, we generate interpolated SEDs as was done in \cite{shah_predictions_2024,  shah25a}. For further details on the interpolation and generation of synthetic KNe SEDs, see \cite{shah_predictions_2024}. 

\subsection{Afterglow SEDs} \label{sec:afterglow_sed}

To account for the emission from the AG in addition to the KN, we use \texttt{afterglowpy}~\citep{ryan_gamma-ray_2020} to compute our AGs. This package uses semi-analytical solutions to promptly compute the flux-density in a wide-range of physical scenarios.  
\texttt{afterglowpy} is designed for prompt computation of AGs from structured GRB jets, or those with a non-uniform energy profile with respect to the jet axis. Based on observations of GRB 1708017A and the associated AG, it became clear that the jets can have a non-uniform, or tophat, structure \citep{wu_constraining_2018, wu_gw170817_2019}.  Several other structured jets have been identified, such as GRBs 150101B \citep{Troja2018_grb150101b}, 160625B \citep{Cunningham2020_grb16}, and 250704B \citep{swain2025_grb250704b}.  Given the diversity of profiles, we elect to use a Gaussian profile for the jet models in this study, which encodes a smoothly varying structured jet.  
Given the nature of jets and their internal structure, there is an intrinsic viewing angle dependence when observing AGs, but the appearance depends on several other factors, as well \citep{Sari_1998_afterglow}. This includes the isotropic-equivalent energy along the jet axis $E_0$, opening angle of the jet $\theta_{\rm core}$, and the density of the medium it is interacting with $n_0$. In the case of GRBs from massive stars, the medium can include material ejected by stellar winds from the massive progenitors. Neutron stars lack significant stellar winds, so the dominantly shocked medium is typically a uniform interstellar medium \cite[ISM; e.g.][]{Chevalier_1999_ism, Li_2020lGRBcsm}. The following microphysical parameters also contribute to the AG's appearance: $p$, the spectral index of the power-law distribution of electron energies; $\epsilon_{\rm e}$, the fraction of shock energy in electrons; and $\epsilon_{\rm B}$, the fraction of shock energy in the magnetic field.  All of these are considered in \texttt{afterglowpy}, along with an additional parameter, the truncation angle $\theta_{\rm trunc}$, or the viewing angle with which the emission goes to zero.

The semi-analytical \texttt{afterglowpy} package is flexible and allows for the flux density to be obtained at any choice of phase and wavelength; however,
the KN SED is limited in its extent in wavelength space, and thus limits where we can include both components. As such, we construct an AG SED with the same grid as the KN SED. In the case of the KN model, the SEDs are computed at intervals of $0.2 \ \rm d$ in the range $t \in [0.1, 19.9] \ \rm d$ \citep{dietrich_multi-messenger_2020} along the time axis. Along the wavelength axis, the SED is computed at every $200$~\AA\  in the range $ \lambda \in [100, 99900]$ \AA. It is possible to evaluate at different points in the phase-wavelength space; however, the interpolator is most consistent when evaluating at points consistent with the original grids.

Table \ref{table:aft-parameters} summarizes the parameters required by \texttt{afterglowpy} along with our selection of values for those parameters, which is explained further in Section \ref{sec:afterglow_param}.  We also note that unless otherwise specified, we use the default \texttt{afterglowpy} configurations.  This does mean that the AGs used in this work will lack a coasting phase; however, for the timescales of interest ($t > 0.1 \,d$), coasting has already concluded.

\section{Simulating Events} \label{sec:simulate}

To better understand the general expectation for coincident KN and AG emission, we randomly generate 50,000 merger events. This number of simulated events allows us to sufficiently sample the distributions presented below.  In simulating these events, it is important that the simulation parameters are chosen to match the population of observed kilonovae. However, many of these bounds are not well known given the small observed sample size, so we conservatively adopt large ranges that are likely to encompass the physically plausible bounds. Recent studies suggest that the fraction of NSMs that launch successful jets is significant, tending toward unity \citep{beniamini_lesson_2019, Sarin_2022_linkingGRBs}. Given this, we assume that every event will have both a KN and AG.  Determining the correlations between KNe and AGs is still an active area of research \citep[e.g.][]{Rastinejad_2024_kngrbs} and thus joint distributions of the physical properties are not well-known, so we elect to sample parameters for any given event's KN and AG independently.  Extrinsic parameters such as the viewing angle, position on the sky, Galactic extinction, and host extinction are shared between the two.  The AG's peak emission and brightness are strongly related to viewing angle, so we generate events viewed within $30 \ \rm degrees$ of the polar axis. This is roughly twice the largest opening angle for a majority of the GRBs with measured opening angles \citep{rouco_escorial_jet_2023}. 

\subsection{Kilonovae} \label{sec:kn-parameters}

For the simulation of the KNe, we use the same parameter distributions as in \cite{shah_predictions_2024}, other than viewing angle, as shown in Table \ref{table:KN-parameters}. Below, we describe our choices in parameter selection.

Of the two neutron star mass distributions typically considered for NSMs, we elected to sample from the \cite{Galaudage_2021_nsmass} mass distribution over \cite{Farrow_2019bnsmass}.  The \cite{Galaudage_2021_nsmass} distribution captures both AT2017gfo \citep{shah_predictions_2024} and the estimated ejected masses from the KN associated with GRB230307A \citep{Bulla_2023_grb23}, which are $m_{\rm ej}^{\rm dyn} = 0.005$ and $ m_{\rm ej}^{\rm wind} = 0.05$. Additionally, for GW190425, an event debated to be either a NSM or a black hole-neutron star merger \citep{Abbott_2020_gw190425}, 
\cite{Galaudage_2021_nsmass} is able to explain the measured masses.  To get the mass of ejected material, we need to select an Equation of State (EOS) for the neutron stars \citep[e.g.][]{Lattimer_2007, Hotokezaka_2011_ejecta}, which parametrizes the relationship between density and pressure and, in turn, the possible mass and compactness of neutron stars. Following the prescription in \cite{shah_predictions_2024}, we use the SFHo neutron star EOS \citep{Steiner_2013eos} with the ejecta fitting functions described in \cite{Setzer_2023ejf} to calculate the resulting ejecta masses from sampled neutron star binaries.

\begin{deluxetable*}{lll}
    \tablecaption{Kilonova Simulation Parameters \label{table:KN-parameters}}

    \tablehead{
        \colhead{Parameter} & \colhead{Description} & \colhead{Values}
    }

    \startdata
    \toprule
    $m_{\rm ej}^{\rm dyn}$, $m_{\rm ej}^{\rm wind}$ & Dynamical and disk wind ejecta masses & see Section \ref{sec:kn-parameters} \\
    $\Phi$ &  Opening angle of the lanthanide region & $\mathrm{U}(15^{\circ}, 75^{\circ})$ \\
    $\cos \theta_{\rm view}$ &  Cosine of the viewing angle & $ \mathrm{U}(\cos 30^{\circ}, 1)$ \\
    $A_V$ &  Host extinction & $\exp \left(-\frac{A_V}{ \tau_V}\right)$ \\
    \enddata
    
\end{deluxetable*}

To remain within reasonable limits of the KN SED interpolator, we sample $\Phi$ uniformly from $[15, 75]$, as in \citep{shah_predictions_2024}. Values of $\Phi$ beyond this region, require extrapolation due to the limits of the grids used in the interpolation.

The events were given a random coordinate on the sky, as locations of host galaxies should have no spatial dependence. We then use the sky position to obtain an associated Galactic reddening from the \cite{1998_Schlegel_dust} dust map.  For the host extinction, we sample from an exponential distribution in the form 
\begin{equation}
    P(A_{\rm V}) = \exp \left(-\frac{A_V}{ \tau_V}\right)
\end{equation} 
where $\tau_{\rm V} = 0.334$,
as found for a sample of extra-galactic supernovae \citep{2009_Kessler_Av}.  The reddening is then found by $E(B-V) = A_{\rm V} / R_{\rm V}$, where $R_{\rm V} = 3.1$.  Both host and Galactic extinction are applied to the SEDs using \texttt{sncosmo}'s F99 \citep{F99_extinction} extinction model. All of the previously discussed parameter selections are summarized in Table \ref{table:KN-parameters}.

\subsection{Afterglows} \label{sec:afterglow_param}

For each simulated KN, we also simulate an AG. Since the KN and GRB are sourced from the same progenitor, we take the viewing angle from the KN to be that of the AG. All other parameters are selected from distributions in recent literature as described below and summarized in Table \ref{table:aft-parameters}. 

To obtain a jet opening angle, we sample from the combined posterior distribution of 10 GRBs with measured opening angles from \cite{rouco_escorial_jet_2023}. They note that the region of non-zero probability for $\theta_c > 15^{\circ}$ is due to just two events, thus we sample from the first peak in the distribution, truncating at $\theta_{\rm c} = 12.4^{\circ}
$ (see Figure 4 in \citet{rouco_escorial_jet_2023}).  We take the truncation angle $\theta_{\rm trunc}$, or angle at which the energy goes to zero, to be $10 \theta_{\rm c}$ as was done in \cite{ryan_gamma-ray_2020, Sarin_2022_linkingGRBs}.

\begin{figure}
{    \includegraphics[width=0.5\textwidth]{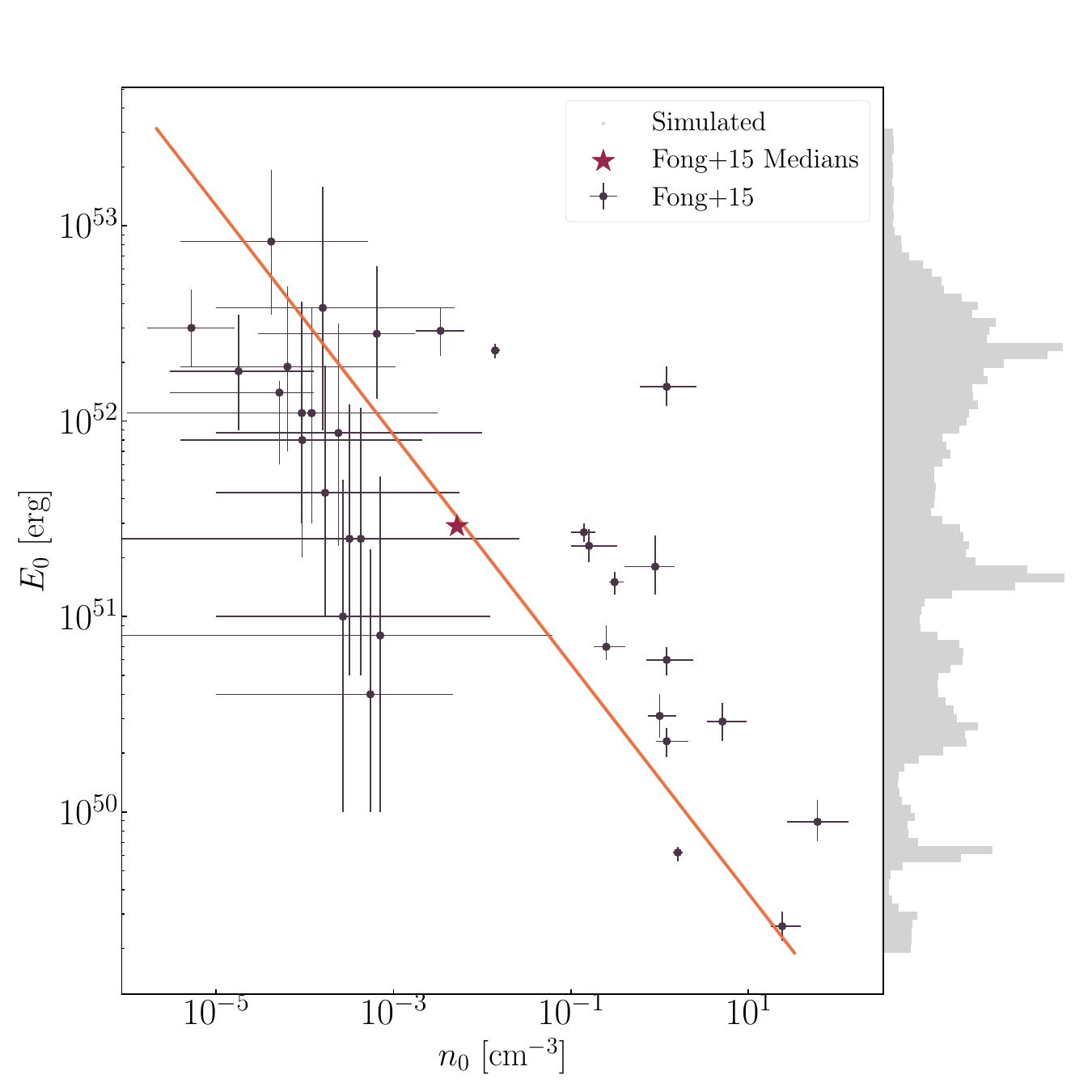}
  \caption{The parameter space of $E_0$ and $n_0$ as covered by the sample of afterglows reported in \cite{fong_decade_2015} (dark red points).  The red star marks the location of the median values reported for the sample in Table 4 of \cite{fong_decade_2015}.  The right and upper histograms show this distribution of $E_0$ values of the simulated sample of afterglows.  Since $n_0$ is determined directly from the sampled distribution of $E_0$ and the fit to the \cite{fong_decade_2015} data (orange line), the distribution is identical, so it has been omitted.}
    \label{fig:E0n0}
}
\end{figure}

As stated in \cite{fong_decade_2015}, constraining $\epsilon_{\rm e}$ and $\epsilon_{\rm b}$ requires well-sampled multi-wavelength photometry, from X-ray to radio, to identify locations of three break frequencies. As a result, $\epsilon_{\rm e}$ and $\epsilon_{\rm b}$ are often fixed when deriving other physical parameters from observed AGs. To be consistent with reported values of observed GRB AGs, we fix $\epsilon_e = 0.1$ and $\epsilon_B = 0.01$. This choice is consistent with known $\epsilon_e$ constraints \citep{beniamini_electrons_2017, duncan_constraints_2023}, and it is often found that $\epsilon_B < 0.1$ \citep[e.g.][]{BarniolDuran_2014_eb}. 

Additionally, it is shown that the choice of $\epsilon_{\rm e}$ influences the inferred $n_0$ and $E_0$.  Through the models and observations from which these two parameters are inferred \citep{fong_decade_2015}, they are degenerate.  As $E_0$ is related to the nature of the binary system and merger and $n_0$ relates to the environment of the burst \citep[which is not necessarily the birthplace of the binary system as they tend to migrate, e.g.][]{Belczynski1999_natalkick, Giacobbo2018_mergerenvs, Mandhai_2022}, they are not necessarily physically correlated. However, allowing these two parameters to vary independently yields combinations of $E_0$ and $n_0$ which lie far from the observed population.  To account for this modeling degeneracy, we elect to model the parameters as being correlated in the following way. First, we compute a simple linear fit, as shown in Figure \ref{fig:E0n0}, to the subsample from \cite{fong_decade_2015}, where $E_0$ and $n_0$ were inferred using $\epsilon_B = 0.01$.  We, then, select the appropriate \cite{fong_decade_2015} cumulative distributions function (CDF) of $E_0$ for our choices of $\epsilon_{\rm e}$ and $\epsilon_{\rm b}$.  To obtain values of $E_0$, we perform inverse transform sampling. Then with these values of $E_0$, we obtain a corresponding $n_0$ from the linear fit.  We elected to sample from the $E_0$ CDF, as the sample has greater relative uncertainty in $n_0$ than in $E_0$. 

We also use the \cite{fong_decade_2015} distribution of the spectral index $p$. As a note, despite one of the AGs in the \cite{fong_decade_2015} sample having $p = 1.92$, we assert that $p > 2$, as is required by \texttt{afterglowpy} to avoid a divergent total energy for the accelerated electrons.  For the fraction of accelerated electrons, we opt for the default parameter choice in \texttt{afterglowpy}, $\xi = 1$, which is within recent constraints \citep{duncan_constraints_2023}. It does seem reasonable to believe $\xi$ to be less than 1; however, similar to $\epsilon_e$ and $\epsilon_B$, it is often fixed as constraining other parameters (e.g. $E_0$ and $n_0$) from observed AGs takes precedence.

\begin{deluxetable*}{lll}

    \tablecaption{Afterglow SED Parameter Distributions \label{table:aft-parameters}}

    \tablehead{
    \colhead{Parameter} & \colhead{Description} & \colhead{Values}
    }
    
    \startdata
    \multicolumn{3}{c}{\textit{Distribution}}\\
    \hline
    $\log E_0$ & log On-axis Isotropic Equivalent Energy & \cite{fong_decade_2015} \\
    $\theta_c$ & Opening angle of the jet core & \cite{rouco_escorial_jet_2023} \\
    $\cos \theta_{\rm view}$ & Cosine of the viewing angle & $\mathrm{U}(\sqrt{3}/2, 1)$ rad \\
    $p$ & Power-law index of electron energies & \cite{fong_decade_2015} \\
    $\log n_0$ & log Density of interacting medium & \cite{fong_decade_2015}\tablenotemark{a} \\
    \hline
    \multicolumn{3}{c}{\textit{Fixed}}\\
    \hline
    $\theta_{\rm trunc}$ & Truncation angle & $\min(10\theta_c,\pi/2)$ \\
    $\epsilon_{\rm e}$ & Fraction of shock energy in electrons & 0.1 \\
    $\epsilon_{\rm B}$ & Fraction of shock energy in magnetic fields & 0.01 \\
    $\xi_{\rm e}$ & Fraction of accelerated electrons & 1 \\
    \enddata
    
    \tablenotetext{a}{The distribution is not directly from \citet{fong_short_2022}. 
    As described in Section~\ref{sec:afterglow_param}, we sample from the $E_0$ 
    distribution and determine $n_0$ from a linear model fit to the measured values 
    of $\log E_0$ and $\log n_0$ for the \citet{fong_short_2022} sample of afterglows.}

\end{deluxetable*}

\section{Synthetic Photometry of the Combined Kilonova and Afterglow Emission} \label{sec:combinedLightcurves}

\begin{figure*}
    \centering
    \includegraphics[width=\textwidth, scale=0.75]{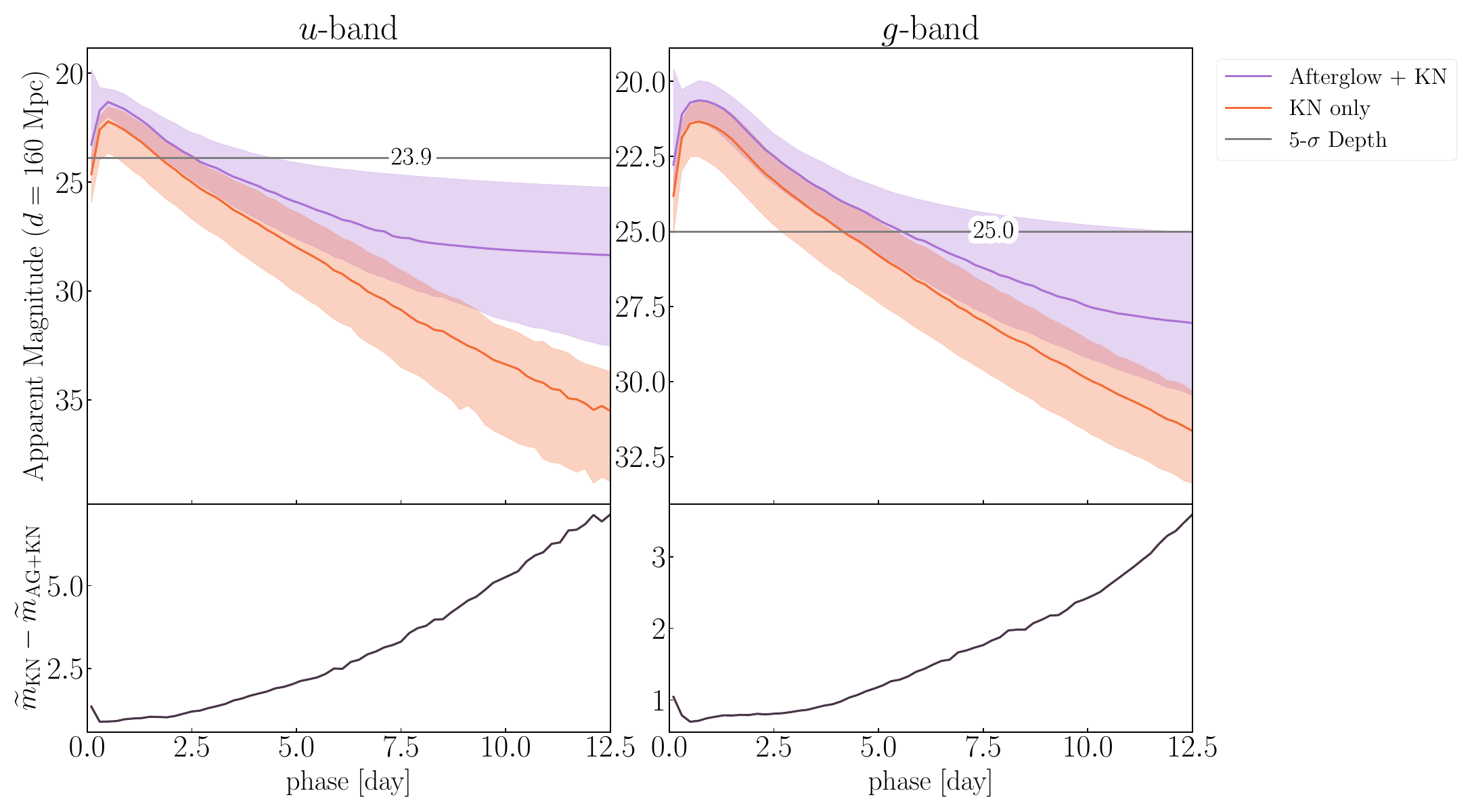}
    \caption{KN (orange) and KN and AG (blue) median light curves $u$ and $g$-bands LSST bands for events at a distance of $160 \ \rm Mpc$, corresponding to the LIGO range of NSMs with two $1.4 \ \rm M_{\odot}$ neutron stars. The respective transparent regions represent the area between the 16th and 84th percentiles and the LSST 5-$\sigma$ depths are indicated with the gray lines. $rizy$-band light curves are omitted here due to the lack of significant enhancement from the AG.}
    \label{fig:lsst}
\end{figure*}

With the combined emission from the KNe and AGs of 50000 events, we then use \texttt{sncosmo} \citep{sncosmo} to obtain the synthetic photometry.  This package integrates the simulated SEDs over the requested passbands, with native support for the LSST bands, while also applying host galaxy and Milky Way extinction.  Here, if necessary, we could elect to include a redshift-dependent $k$-correction; however, we are only interested in distances for which the KN emission in observable ($\lesssim 600\, \rm Mpc$). With LSST depths this is still $z \simeq 0$, so we assume it to be negligible. Figure \ref{fig:lsst} shows the median, 16th and 84th percentile light curves relative to the 5-$\sigma$ depths of 23.9 and 25.0 in $u$- and $g$-bands, respectively \citep{Bianco_2022_optobs}.  These percentiles are taken per time step of the simulated KNe, such that the region bounds typical light curves of the simulated sample.  In the $u$-band, we see the greatest deviation from the KN-only.  While the greatest enhancement occurs at phases when the event is likely to be faint ($t > 5 \ \rm day$), even for LSST, the difference in the median curves prior to 5 days is $>\! 0.8 \ \rm mag$.  In the $g$-band, where LSST is more sensitive, the difference in medians prior to 5 days is $>\! 0.6 \ \rm mag$.  In Appendices \ref{fig:uvex} and \ref{fig:jwst}, we also consider the combined emission in the ultraviolet and infrared with the {\it UltraViolet EXplorer} ({\it UVEX}) and the {\it James Webb Space Telescope}, respectively.  The limited sensitivity of {\it UVEX} and the lack of significant contributions from the AG in the infrared made these photometric regimes less relevant to this work.

\begin{figure}
    \includegraphics[width=0.45\textwidth]{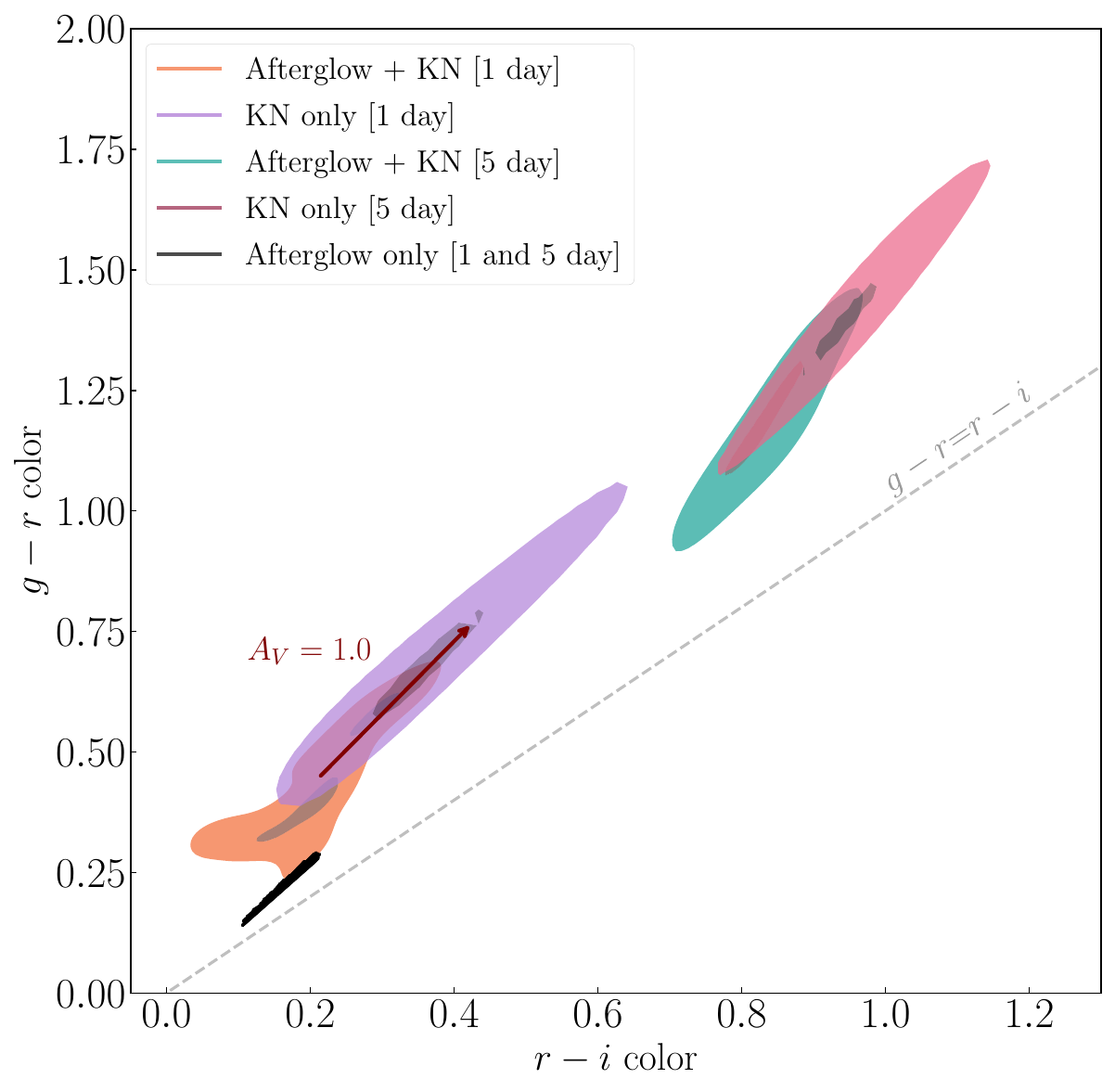}
    \centering
    \caption{The $68\%$ and $95\%$ contours of $g-r$ color vs $r-i$ color of only afterglow (black), kilonova 1-day post-merger (purple), and kilonova 5-days post-merger (red) emission. Along with the combined emission at 1 (orange) and 5 days (teal) post-merger.  The afterglow evolution is negligible compared to that of the KNe and thus only occupies a line in this color-color space.  For the combined emission at later times, there is a small region separate from the majority due to combined events which are dominated by AG emission, rather than KN emission, leading to the contour appearing to stretch between the regions of the color space where the AG only and KN only lie; however, for the majority of events the color-color information is able to separate most events for which there is KN emission present.  Additionally, a reddening vector (red arrow) shows the direction a GW170817-like KN \citep{dietrich_multi-messenger_2020} at 1 day post-merge would move starting at $A_V=0$ to $A_V=1$.}
    \label{fig:colorcolor}
\end{figure}

It is also interesting to consider the color of these combined events.  \citet{zhu_kilonova_2022} finds that the color of combined KN and AG emission can be used to distinguish them from other transient events. Figure \ref{fig:colorcolor} shows the $g-r$ vs. $r-i$ at a phase of 1 and 5 days post merger for the combined events as well as only the KNe and only the AGs to represent events for which one component is clearly dominant and appears as if it were the only source of emission.  A similar color evolution exists for other permutations of LSST color; however, the one presented in Figure \ref{fig:colorcolor} acted as the middle ground for showing that the AG does pull the distribution to zero color while still showing that the combined emission is distinct.  The redder colors have nearly all KN emission where the bluer colors had notable AG contamination.   Also, in phase, from 1 day post-merge to 5 days.  For clarity, in phase, we show only 1 and 5 days post merge, but there is a  there is an evolution from near zero towards 1 mag.  The combined emission in this color-color space has events for which the AG or KN dominates, which contributes to the appearance of being stretched to both extremes.  From this, we can see that this color information can be used to distinguish events with KN emission from those that dominated by the afterglow.  Events for which the AG dominates, there is negligible color evolution; however, when the KN contributes, there is a clear color $>\! 0.5 \ \rm mag$ at 5 days post merger. Such color analysis has been used to identify two KNe coincident of long GRBs, as a red excess was present in both \citep{Rastinejad_2022_grb211211, Bulla_2023_grb23, Levan_2023_grb23}.  Additionally, in comparison with simulated LSST transients from the "Extended LSST Astronomical Time-series Classification Challenge" \citep[ELAsTiCC \footnote{\url{https://portal.nersc.gov/cfs/lsst/DESC_TD_PUBLIC/ELASTICC/}};][]{elasticc}, we find that other short-lived transients (e.g. M-dwarf flares, dwarf novae, and microlensing events) that may be contaminants or sources of confusion with NSM emission do not traverse the color-color space in a similar way.

It is also important that we consider the effect of dust on the color evolution.  To do so, we first simulate an on-axis GW170817-like KN using the median parameters from the \citet{dietrich_multi-messenger_2020} analysis of AT2017gfo ($\log m_{\rm ej}^{\rm dyn} = 10^{-2.27}, \log m_{\rm ej}^{\rm wind} = 10^{-1.28}, \Phi=49.5$).  We then obtain the $g-r$ and $r-i$ colors at 1 day post-merger for $A_V = 0$ and $A_V = 1.0$ to determine the direction extinction would move a point in the $g-r$-$r-i$ space.  This vector is shown in Figure \ref{fig:colorcolor}.  It is possible for dust to redden the event in a manner similar to that of its natural color evolution. While $A_V \gtrsim 1$ has been observed for some short GRBs hosts \citep{nugent_short_2022}, this level of extinction would likely suppress the intrinsically faint KN signal below detection limits.  On the other hand, in order to see the AG along with the KN, the jet would need to be oriented toward the observer, and thus the extinction is only from the line-of-sight through the host and the Milky Way.

\section{Afterglow Scalings} \label{sec:scaling}

By following the analytical descriptions of AGs by \cite{Sari_1998_aftLC} and \cite{granot_shape_2002}, we can understand how the photometric evolution of the AGs, as seen in Figure \ref{fig:lsst}, scales with the relevant physical parameters over time. For the wavelength range of interest for this work, \cite{Sari_1998_aftLC} shows that synchrotron self absorption is not important and thus we can follow the same prescriptions which describes two main regimes of an AG's evolution: radiative and adiabatic cooling of the shock.  When the fraction of the shock energy in electrons is large, i.e. $\epsilon_{e} \rightarrow 1$, the evolution begins radiative and transitions to adiabatic at a time $t_0$: 
\begin{eqnarray}
    t_0 & = & 4.6 \ \epsilon_{B}^{7/5}\epsilon_{e}^{7/5} \left( \frac{E_0}{10^{52} \ \rm erg}\right)^{4/5} \left( \frac{\gamma_0}{100}\right)^{-4/5} \nonumber \\
   & & \times \left( \frac{n_0}{1 \ \rm cm^{-3}}\right)^{4/5} \, \rm days
\end{eqnarray}
where $\epsilon_{B}$ is the fraction of shock energy in magnetic fields, $E_0$ is the on-axis isotropic equivalent energy, $\gamma_0$ is the initial Lorentz factor of the material ejected in the merger, and $n_0$ is the density of the interaction medium \citep{Sari_1998_aftLC}.  Given the default value from \texttt{afterglowpy} of $\gamma_0 = \infty$, the transition occur immediately, and the adiabatic evolution can describe the AGs.  Then, via Equations (11) of \cite{Sari_1998_aftLC}, we find that for the observed frequencies ($\nu_u \approx 8 \cdot 10^{14} \ \rm Hz$ and $\nu_g \approx 6 \cdot 10^{14} \ \rm Hz$ for $u$- and $g$-bands, respectively) and times of interest ($t \lesssim 20 \ \rm days$), $\nu_m < \nu < \nu_c$ indicating that the \cite{granot_shape_2002} `G' power-law segment captures the AG evolution, aligning with typical assumptions for optical afterglows. The flux density of this segment is given by:
\begin{eqnarray}
        F_{\nu, \rm G} & = & 0.461 \, (p-0.04)\, e^{2.53p}\, (1+z)^{(3+p)/4}  
        \nonumber \\
        & \times &   \epsilon_e^{-p-1} \, \epsilon_B^{(1+p)/4}  
        \pfrac{n_0}{1 \ \rm cm^{-3}}^{1/2} \pfrac{E_0}{10^{52} \ \rm erg}^{(3+p)/4}  
        \nonumber \\ 
        & \times & \pfrac{t}{1 \ \rm day}^{3(1-p)/4} \pfrac{d_L}{10^{28} \ \rm cm}^{-2} \pfrac{\nu}{10^{14} \ \rm Hz}^{(1-p)/2}
        \label{eq:AGflux}
\end{eqnarray}
where $p$ is the power-law index of the electron energies.  For the distances considered here, $1+z \approx 1$, and ignoring constants, the absolute magnitude can be approximated as $M \approx -\log_{10}(F_{\nu})$. As seen in Figure \ref{fig:lsst}, the KNe evolves as $dm/dt \sim 1 \ \rm mag/day$ and the AG evolves much more slowly than the KN, giving rise to the plateau in the light curves at $t > 10 \ \rm d$ where the AG begins to dominate. 

Equation (\ref{eq:AGflux}) sets the afterglow flux and luminosity and so will control our predictions.  To anticipate results below, a typical value of $p \sim 2.2$ gives a scaling
 $F_{\nu} \propto n_0^{0.5} E_0^{1.3}$ and given the range of possible $p$ values, the scaling ranges from $F_{\nu} \propto n_0^{0.5} E_0^{1.25}$ to $F_{\nu} \propto n_0^{0.5} E_0^{1.5}$.
We see that the afterglow energy has a very strong scaling, while ambient density $n_0$ has a much weaker scaling. Thus, the assumed range of $E_0$ values will play the strongest role in our predictions below, but $n_0$ will also be important.  Also, because our results depend on the product, our results will be sensitive to correlations between these variables in the data we use, as described in Section \ref{sec:afterglow_param}.

\section{Discovery Rates} \label{sec:discovery}

\subsection{Discovering Neutron Star Mergers with LSST} 

To assess discovery with LSST, we conduct the following analysis on the simulated sample.  We consider a first successful observation at time $t_0$, relative to merger, to be when the apparent magnitude exceeds the limiting flux in any two bands at $t_0$ or in the next adjacent time bin, $t_0+0.2 ~ \rm d$.  This acts as a proxy for LSST intra-night gaps, which is the time between when prompt revisits of a point on the sky, on the order of hours\footnote{\url{https://usdf-maf.slac.stanford.edu/allMetricResults?runId=2\#IntraNight}}.  To further consider this event for discovery, the 1st successful observation must also have been preceded by a previous unsuccessful observation at $t_{\rm prev} = t_0 - \delta t_{\rm inter}$, where $\delta t_{\rm inter, prev}$ is the inter-night gap, or time between LSST visits on the order of days\footnote{\url{https://usdf-maf.slac.stanford.edu/allMetricResults?runId=2\#InterNight}}.  Additionally, the event is detected and recovered if there is a successful observation at $t_{\rm next } = t+\delta t_{\rm inter, next}$, with at least one band in common with the observation at $t_0$. 

\begin{figure*}
    \centering    \includegraphics[width=\textwidth, scale=0.75]{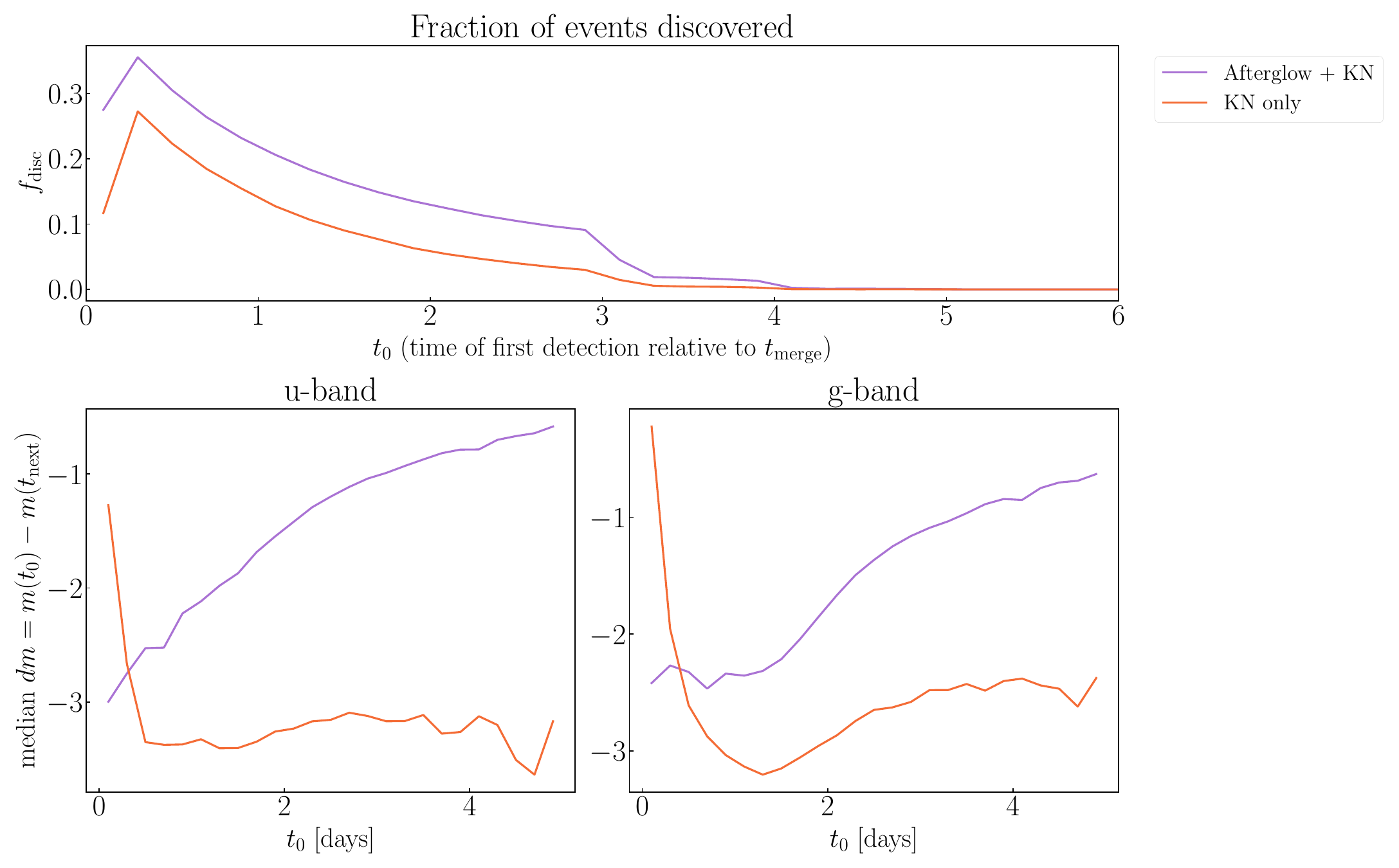}
    \caption{{\it Top}: The fractions of simulated KN and combined events detected in two LSST observations as a function of time of 1st observation relative to time of merger, $t_0$. {\it Bottom}: The median change in observed magnitude from 1st to 2nd observation in $u$-band (right) and $g$-band (left).}
    \label{fig:eff_dm}
\end{figure*}

For the simulated sample, we select a fixed $t_0$ from the phases.  We select $t_0$ at $0.4 ~d$ intervals from the phases at which the models are evaluated.  Using the {\tt baseline\_v4.3.5\_10yrs} Metrics Analysis Framework (MAF) simulation 
\citep{peter_yoachim_2026_18331438},
we sample from the distribution of median inter-night gaps from pointings within Wide Fast Deep survey area, excluding the ``bulgy" region in the galactic plane, for each event to get $\delta t_{\rm inter, prev}$ and $\delta t_{\rm inter, next}$.  We do not consider values of $t_0$ in which $t_{\rm inter, next}$ exceeds the maximum phase of the models ($t > 20 \,d$); however, this occurs at phases much later than when the discovered fraction goes to zero. We compute the fraction of the simulated sample that is recovered with LSST at each $t_0$, shown in Figure \ref{fig:eff_dm}.  The additional on-axis AG emission does improve the discovery of events with LSST, relative to KN emission alone, with peak efficiency aligning with peak KN emission.  The sharp features in the curve are a property of how the detection and discovery criteria are defined, as well as the finite sampling of $t_0$. The peak comes from events transitioning from not detected at $t_0$ to being detected but not in the same bands at both $t_0$ and $t_{\rm next}$ to meet our discovery criteria.  This then falls as $t_0$ occurs later in the photometric decline because at $t_{\rm next}$, they have faded below detection.  The feature at $t_0 \simeq 3$ days arises from events that are detected at $t_0$, but are also detected at $t_{\rm prev}$ meaning they should've already been discovered. These are not new and are not counted.  Then lastly, at around $t_0 = 4$ days, events are typically not bright enough to be detected at $t_{\rm next}$.

Additionally, we measure the change in brightness of the event between the first and second observation, as shown in the bottom of Figure \ref{fig:eff_dm}.  Since, as discussed in Section \ref{sec:scaling}, KN $dm/dt \sim 1 ~d$ and the median inter-night gap is $\sim 3 \rm ~ d$, after peak emission the change in KN brightness is also $\sim 3 \rm ~mag$.  For the combined emission, as the first detection occurs later in the event evolution, the AG emission reduces how much the event fades. This occurs most strongly in $u$-band, as one might expect from the median light curves shown in Figure \ref{fig:lsst}, and again the AG contributions in the bands redder than $g$ are not as significant and thus are not shown. 

\begin{figure*}
    \centering    \includegraphics[width=\textwidth, scale=0.75]{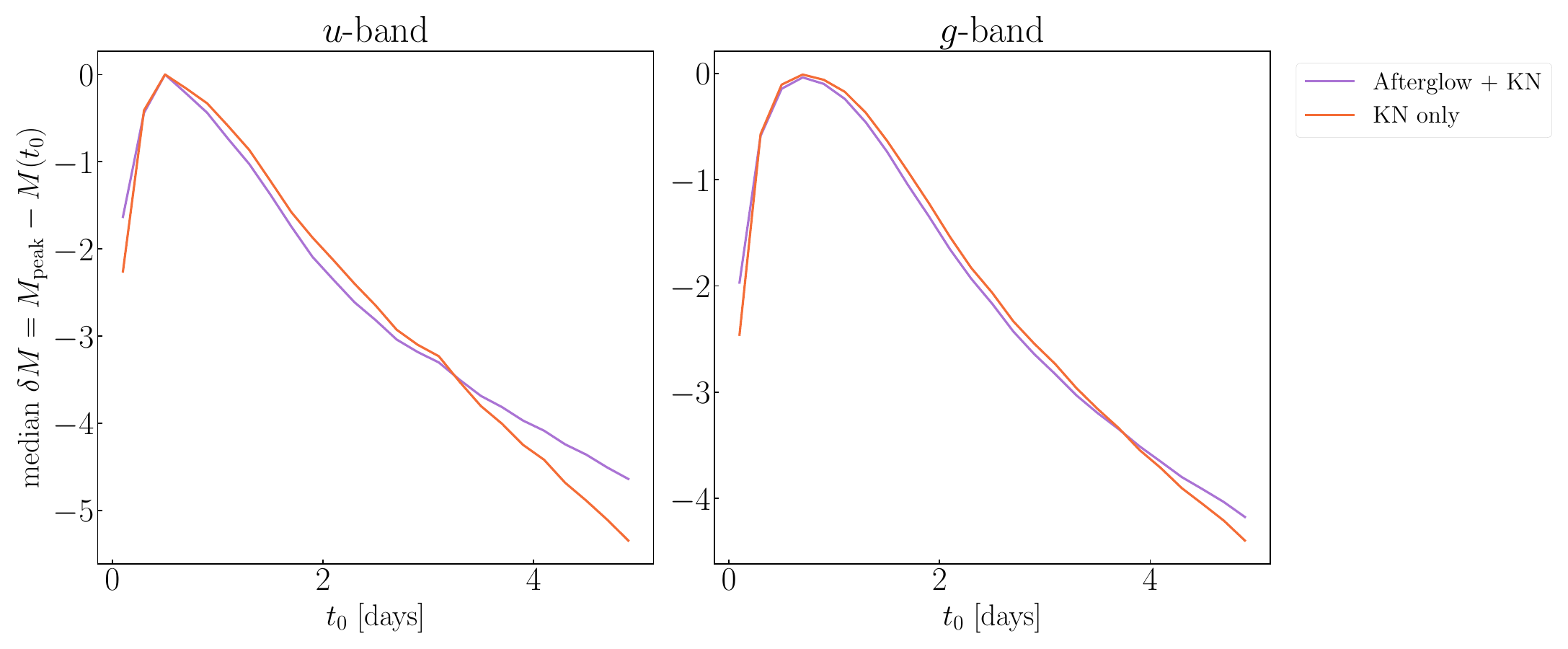}
    \caption{The median change in absolute magnitude from peak to 1st observation in $u$-band (right) and $g$-band (left) over time of first observation.}
    \label{fig:abs_dm}
\end{figure*}

\subsection{Luminosity Function and Rates} \label{sec:rates}

Suppose the absolute magnitude of the event is fixed, the maximum distance $r_{\rm max}$, in $\rm pc$, for which the event could be detected with limiting magnitude $m_{\rm lim}$ is 
\begin{equation}
     r_{\rm max}(M, m_{\rm lim}) = 10^{\frac{1}{5}(m_{\rm lim}-M) + 1}
\end{equation}
and thus the observable volume for a telescope with a sky coverage of $f_{\rm obs} = \Omega_{\rm obs}/ 4\pi$ is that of a sphere of radius $r_{\rm max}$,
\begin{eqnarray}
	V(M, m_{\rm lim}) &=\frac{4\pi}{3} f_{\rm obs} \cdot 10^{\frac{3}{5}(m_{\rm lim}-M) + 3}  \nonumber \, .
\end{eqnarray}
From this, we can see a strong dependence on the peak magnitude ($ \propto 10^{-3M/5}$).
Then the rate $\Gamma$ is 
\begin{equation}
    \Gamma(M, m_{\rm lim}) = \volrate \cdot V(M, m_{\rm lim}) = \volrate f_{\rm obs} \cdot 10^{\frac{3}{5}(m_{\rm lim}-M) + 3}
\end{equation}
where $\volrate$ is the volumetric rate of events. For a distribution of events, the average rate is 
\begin{eqnarray} \label{eqn:rate_peak}
    \left< \Gamma(M, m_{\rm lim})\right> & = & \volrate \cdot \left<  V(M, m_{\rm lim})\right> \\
	& = & \volrate f_{\rm obs} \cdot 10^{\frac{3}{5}m_{\rm lim} + 3} \cdot \left<10^{-3M/5}\right> \nonumber
\end{eqnarray}
where
\begin{equation}
    \left <10^{-3M/5} \right> = \frac{\int \phi(M) 10^{-3M/5} dM }{\int \phi(M) dM }
\end{equation}
and $\phi(M)$ is the distribution of absolute magnitudes, i.e., the (peak) luminosity function. 

\begin{figure*}
    \centering    \includegraphics[width=\textwidth, scale=0.75]{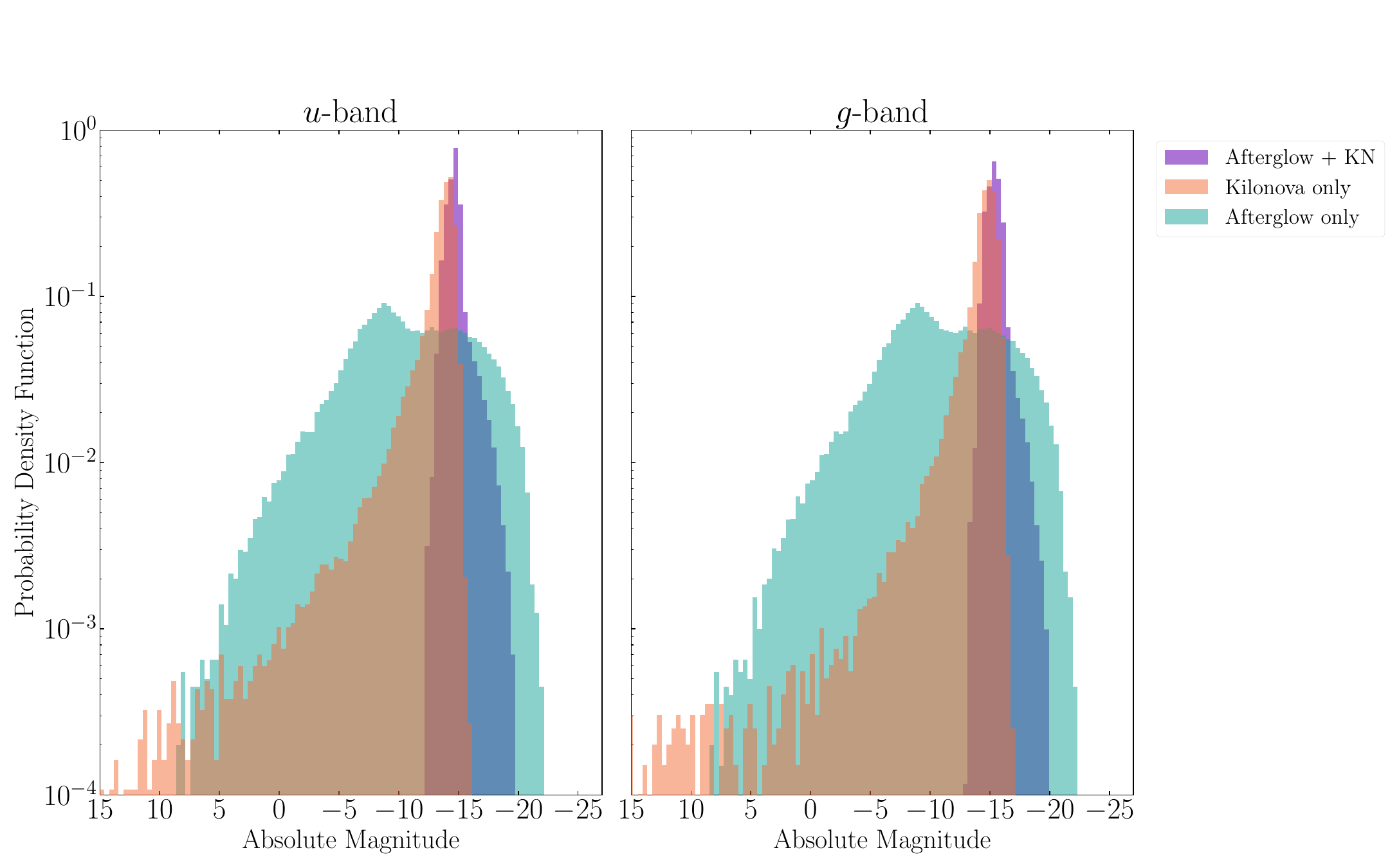}
    \caption{The distributions of peak $u-$ and $g-$band magnitudes for the simulated sample of KN, AGs, and the combined emission with a cut at $g-r$ and $r-i > \rm 0.5 \ mag$.}
    \label{fig:lumFunc}
\end{figure*}

We find the distribution of peak magnitudes, shown in Figure \ref{fig:lumFunc}, from which we compute the rate of observable events with LSST.  As expected, the KNe have a relatively narrow range of peak magnitudes relative to that of the AGs.  Relative to a the distribution fo simulated AGs from \cite{zhu_kilonova_2022}, our sample has events brighter than -20 in absolute magnitude and is a result of the different choices of AG parameters.  Prior to computing the rate of combined events, we make a cut to remove events dominated by AGs to focus on KNe enhanced, but not hidden by the AG.  As seen in Figure \ref{fig:colorcolor}, a majority of events containing KN emission become distinct in the $g-r$-$r-i$ color space from AG only events only a few days post-merge.  We make a conservative cut at $g-r > 0.5 \ \rm mag$ and $r-i > 0.5 \ \rm mag$ to remove the combined events for which the AG is most dominant.  This can be seen in the difference between the AG only and total peak magnitude distributions in Figure \ref{fig:lumFunc}.  Additionally, by making this cut to remove bright AGs, the redshift evolution of the NSM rate becomes negligible, as we are only considering events for which the KN is observable, which even with LSST depths is $z \simeq  0$.  The resulting luminosity function also consistent with results in \citet{zhu_kilonova_2022}. 

We would like to note that the distribution of AG peak magnitudes does include a few very bright events ($M_g \lesssim -20$) and this is a product of our method of simulating, as described in \ref{sec:afterglow_sed} and \ref{sec:afterglow_param}.  When comparing the simulated sample luminosities to that of the observed sample from \citet{Castrejon2025_hostafterglows}, the upper limits are comparable and the simulated sample has a much smaller median value, so our simulated sample is not beyond what has been observed.  However, these bright events however will dominate a rate computed with the method described below in a way that yields unreasonable large results, and so we do not report a AG-only rate in this work.

In order to get an estimate of the uncertainty in the effective volume from the simulated events, we take 20 subsamples of $5000$ events to compute the median volume and take the $1-\sigma$ errors as the uncertainty on the volume used to determine the rate.
Next, we must select an estimate of the intrinsic volumetric rate of NSMs\footnote{A flat $\Lambda$-CDM cosmology is assumed such that $h_{70} = h/0.7 = 1$. All volumetric rates discussed in this work have been scaled by the appropriate $h_{70}^3$, as needed.}. The rate of NSM has been measured by several groups. \cite{Abbott_2023gwtc3} estimates the rate to be  $11-1863 \ \rm Gpc^{-3} yr^{-1}$ from the third Gravitational Wave Transient Catalog and \cite{nitz_4-ogc_2023} obtains an estimate of $220^{+340}_{-163} \ \rm Gpc^{-3} yr^{-1}$ from the forth Open Gravitational Wave Catalog.  From binary neutron star systems within the Galaxy, \cite{Chruslinska_2018dns} found a rate of $48 \ \rm Gpc^{-3} yr^{-1}$, which did not align with the current LVK rate at that time, $1540^{+3200}_{-1220} \ \rm Gpc^{-3} yr^{-1}$ \citep{Abbott_2017_gw}, but is consistent with the latest estimate of $56^{+99}_{-40} \ \rm Gpc^{-3} yr^{-1}$ from candidates sourced from the Gravitational-Wave Candidate
Event Database (GraceDB) alert stream \citep{Akyuz2025rate}.

When selecting the \cite{Akyuz2025rate} volumetric rate and optimistic fractional sky coverage of $0.5$ with Equation \ref{eqn:rate_peak} with the distribution of peak magnitudes, the rate of detectable KN with LSST is 
\begin{eqnarray}
    \left< \Gamma^{\rm peak}_{\rm KN, u} \right> &=& \asymuncert{6.8}{12}{5} \ \rm yr^{-1}  \\
    \left< \Gamma^{\rm peak}_{\rm KN, g} \right> &=& \asymuncert{114}{202}{82} \ \rm yr^{-1} \, .
\end{eqnarray}  The uncertainty in the volumetric rate is the most dominant.  We also recover the \citep{Galaudage_2021_nsmass} $r$ band rate of $\asymuncert{2}{3}{2}$ from \citep{shah_predictions_2024} when using the ZTF-$r$ limiting magnitude of $21.4$.  As is often done, we assume that KNe are roughly isotropic; however, AGs have a strong dependence on viewing angle and the simulated sample are viewed within $30 \ \rm deg$ ($\pi / 6 \ \rm rad$) of the polar axis, thus we much correct Equation \ref{eqn:rate_peak} with a factor 
\begin{eqnarray}
    f_{30} &= & \frac{ \Omega_{30 \ \rm deg} }{\Omega_{\rm sky}} \\
           &= & \frac{1}{2}[1 - \cos(\pi / 6)] \nonumber \, .
\end{eqnarray} With this correction, the rate from events with combined KN and AG emission is
\begin{eqnarray}
    \left< \Gamma^{\rm peak}_{\rm total, u} \right> &=& \asymuncert{45}{80}{32} \ \rm yr^{-1} \label{eqn:rate_total_u}\\
    \left< \Gamma^{\rm peak}_{\rm total, g} \right> &=& \asymuncert{361}{639}{256} \ \rm yr^{-1}. \label{eqn:rate_total_g}
\end{eqnarray}
It is important to note that given the strong dependence on peak magnitude the rate is very sensitive to the number of very bright events and thus choice of AG parameters. 

This rate is an overestimate of the observed rate as it assumes the observations were taken at peak; however, this is often not the case.  For an observation at $t_0$, such that $M(t_0) = M_{\rm peak} - \delta M$, the maximum distance for which it is observable is 
\begin{eqnarray}
    r_{\rm max}(M(t_{0}), m_{\rm lim}) & = & 10^{\frac{1}{5}(m_{\rm lim}-M(t_0)) + 1} \\
         & = & 10^{\frac{1}{5}(m_{\rm lim}-M_{\rm peak}+ M_{\rm peak}(t_0)) + 1}  \nonumber \\
         & = &r_{\rm max,peak} \cdot 10^{\delta M/5} \nonumber
\end{eqnarray}
and thus 
\begin{eqnarray}
    \Gamma_{\delta M}& = & \volrate \cdot f_{\rm obs} \frac{4\pi}{3} \left(r_{\rm max,peak} \cdot 10^{\delta M/5}  \right)^3 \label{eqn:rate_nonpeak}\\
                    & = & 10^{3\delta M /5} \cdot \Gamma(M_{\rm peak}, m_{\rm lim}) . \nonumber
\end{eqnarray}
We use this as a means of capturing some loss relative to the perfect observations at peak magnitude.  We assume that the decline in brightness from peak $\delta M$ is independent of the cosmic rate $\Gamma(M_{\rm peak}, m_{\rm lim})$.  However, $\delta M$ depends on the behavior of the light curve through the duration of the decline, and is thus correlated with the cosmic rate. To a first approximation, we take them to be independent such that Equation \ref{eqn:rate_nonpeak} factors as shown. Using Equation \ref{eqn:rate_nonpeak}, in the case of a first observation one magnitude fainter than peak, the rate is instead
\begin{eqnarray}
    \left( \Gamma_{\rm KN, u} \right)_{\delta M = -1} &=& \asymuncert{1.7}{3.0}{1.2} \ \rm yr^{-1} \, , \\ 
    \left( \Gamma_{\rm KN, g} \right)_{\delta M = -1} &=& \asymuncert{29}{51}{21} \ \rm yr^{-1}
\end{eqnarray}
and the rate from the combined emission, again correcting for viewing angle, is
\begin{eqnarray}
    \left( \Gamma_{\rm total, u} \right)_{\delta M = -1} &=& \asymuncert{11}{20}{8.1} \ \rm yr^{-1}  \\
    \left( \Gamma_{\rm total, g} \right)_{\delta M = -1} &=& \asymuncert{91}{160}{65} \ \rm yr^{-1} \, .
\end{eqnarray}

It is useful to compare this rate to that of detectable sGRBs, as they too have a strong angular dependence.  A recent study of sGRB within $200 \ \rm Mpc/h$ estimated the detectable rate to be $1.3^{+1.7}_{-0.8}$/yr, from which we derive a volumetric rate of $\sim 35^{+45}_{-21} \rm Gpc^{-3} yr^{-1}$ \citep{Dichiara_2020grbrate}.  

Additionally, estimates have been made for KNe discovered with a coincident gravitational wave signal. For the duration of the O4 LVK observing run, \cite{shah_predictions_2024} estimated $2^{+3}_{-2}$ events and  \cite{Colombo_2022knrate} suggested $g$-band follow-up will lead to  $5.7^{+8.7}_{-4.2} / \rm yr$ during of O4.  These are consistent with the KN rate corrected for off-peak observations.  The intrinsic rarity of NSMs is difficult to overcome, and to further put it into context how rare these events are, the rate of a common transient, the core collapse supernova, is $9.1^{+1.56}_{-1.27} \times 10^{-5} \ \rm Mpc^{-3} yr^{-1} = 91000 \ \rm Gpc^{-3} yr^{-1}$ \citep{Frohmaier_2020ccsne}.  Despite their rarity, the LSST depths make finding NSMs without gravitational wave context possible.  Since most will be discovered near peak brightness, rapid follow-up can be triggered allowing for confirmation of candidates and eventually a sample of objects to study.

We would like to note further uncertainties present in the rates reported above.  One key uncertainty is the choice of modeling the $E_0-n_0$ parameter space.  As discussed in Section \ref{sec:afterglow_param}, the modeling used to obtain values for these parameters from observed events is degenerate.  Additionally, there is uncertainty in the cosmic rate of mergers.  The values reported here scale linearly with it, but the current rates are poorly constrained.  Similarly, we have assumed here that all mergers successfully launch jets that yield sGRBs; however, there is uncertainty in this as well. 

\section{Summary, Discussion, and Caveats} \label{sec:summary}

Using \texttt{afterglowpy} \citep{ryan_gamma-ray_2020} and interpolated SEDs from \cite{dietrich_multi-messenger_2020} KN grids \citep{shah_predictions_2024}, we investigate the prospects of using afterglows as a means of improving the chances of finding NSMs in LSST. We find that in cases where the events are observed within 30 deg of the polar axis and observed one magnitude fainter than at peak, the rate of discovery in $g-$band is enhanced from $\asymuncert{29}{51}{21} \ \rm yr^{-1}$ to $\asymuncert{91}{160}{65} \ \rm yr^{-1}$.  Thus, it is possible for the AG emission to aid in the discovery of these events.  Additionally, we find that color information, specifically $g-r$ and $r-i$, can help to distinguish afterglows from observations containing KN emission. 

In this work, we assume the physical parameters of the KN and AG are completely independent of one another; however, given the GRB and KN are from the same progenitor system, it seems unlikely this is true in all cases. It has been shown that the material ejected by the merger can collimate the jet as it breaks through \citep[e.g.][]{Ramirez-Ruiz_2002col, Bromberg_2011col, Duffell_2015col, Urrutia_2021col}, and thus there is a correlation between the mass and distribution of ejected material and the structure and opening angle of the jet. This has yet to be parameterized, and as such was not included in this work, but further KN and AGs observations may constrain such correlations or reveal others.  

Similarly, breaking the modeling degeneracy between afterglow energy and the density of the interacting medium will be a critical component for future afterglow studies.  Breaking this degeneracy requires multi-wavelength observations from X-ray to radio to capture each segment of the synchrotron spectrum.  For nearly half of the \cite{fong_decade_2015} sample, the optical and X-ray observations sample the same region of the spectrum; however, for those with additional observations in the radio, the degeneracy is broken, yielding better constraints on $E_{\rm K}$ and $n_0$.  \cite{Laskar_2022} showed that with radio observations that capture a jet break combined with more sophisticated modeling can yield precision measurements of these parameters.  While these radio observations have been shown to be important for studying the physics of afterglows, follow-up programs often conclude too soon \citep{Schroeder_2024_210726A}; however, improved radio observing strategies have resulted in a subset of radio-selected afterglows which have been found to occur in denser mediums than others sGRBs \citep{Schroeder_2025_231117A}.

There are several assumptions made in this work that may not be representative of the real population of NSMs.  
We opt to construct our rate using the volumetric rate of NSMs; however, it may be the case that not all NSMs produce GRBs and therefore afterglows.  As the volumetric rate derived from gravitational wave methods decreases, eventually it will not be possible to explain all sGRBs through NSMs. We also take the existing sample of measured AG parameters, which leads to the likely inclusion of observation biases as the full AG parameter space may not be fully explored in this sample.  As discussed, the \cite{fong_decade_2015} sample used fixed values of $\epsilon_e = 0.1$ and $\epsilon_B = 0.01$, as constraining these values is difficult; however, lower values of $\epsilon_e$ and $\epsilon_B$ yields less luminous afterglows.

Discovery of NSM candidates with LSST can then guide follow-up in the NIR, where the AG is often not as dominant and less likely to obscure the temporal evolution of the KN. All of which will be important in times when gravitational wave detectors are not active and we are relying solely on blind photometric searches.  As showcased with \texttt{ZTFReST} \citep{Andreoni2021_ztfkne}, early time identification of the afterglows enables multi-band follow-up, so it is beneficial to consider LSST in collaboration with other surveys like ZTF \citep{Bellm_2019ztf}, Young Supernova Experiment \citep[YSE;][]{jones_young_2021}, and La Silla Southern Sky Survey \citep[LS4;][]{Miller2025LS4}, all which have overlap with the LSST footprint.  In Section \ref{sec:discovery}, we required two consecutive, successful LSST observations for a discovery; however, this represents the most pessimistic case, as another survey instrument could get the next observation more quickly than the LSST cadence allows for.  While other survey instruments are not as sensitive as LSST, target of opportunity observations with longer exposure times could get to comparable depths to verify candidates and eventually develop a sample of KNe.  We also see the enhancement is greatest in the ultraviolet; however, the single exposure limits are currently not deep enough to capitalize on this enhancement.  Improving these detection thresholds in the ultraviolet would improve the utility of these bands for observing optical AGs, while also better constraining the UV emission from KN.

Identifying NSM candidates will enable host and studies, as currently, there exist sample studies of host galaxy parameters for sGRBs only \citep[e.g. ][]{Fong_2013_hosts, fong_short_2022, nugent_short_2022, Castrejon2025_hostafterglows}.  By developing host population statistics, one can use such information for identifying future event. \cite{Gagliano_2021_ghost} showed that host galaxy information can be used to distinguish between Type Ia and core-collapse supernovae, with limited observed emission from the transient itself. By studying NSM hosts, the host properties can be used as additional context to assess how likely a candidate is of being a true NSM  and thus make discoveries at earlier times.  A similar technique was employed in the discovery of AT2017gfo associated with GW170817 \citep{Coulter_2017S_sss17a}, where the properties of galaxies within the gravitational wave localization region were used to prioritize the follow up with the Swope Supernova Survey that was first to detect the kilonova counterpart.  In the case without gravitational wave information, one could prioritize follow up of candidates which have color evolution \citep{zhu_kilonova_2022} and host galaxy properties consistent with expectation for NSMs.  It also interesting to consider the newly proposed class of GRBs, compact-object GRBs \citep{gottlieb_unified_2023}.  This distinction followed two long-duration GRBs, GRB211211A \citep{Rastinejad_2022_grb211211, troja_nearby_2022} and GRB230307A \citep{Bulla_2023_grb23, Gillanders_2023_grb23, Levan_2023_grb23}, having compact object origins, blurring the line of GRB progenitors the lied between sGRBs from mergers and long GRBs from collapsars.  Thus, developing a sample of NSM hosts is of interest. 

While this work considers events without GW signals, the additional context will also improve the chance of discovery \citep{Saleem_2017_gwgrb}, as the localization maps reduces the search from the entire sky to a smaller area.  The future improvements to the detectors will help to reduce localization areas and may address contention that exists between some estimates of the NSM rates derived from GRBs and GW.  For example, \citet{jin_short_2018} derives a rate from GRBs that is consistent with that of rate from \citet{Abbott_2023gwtc3} and GW but not the latest estimate \citep[e.g.][]{Akyuz2025rate}.

\section{Acknowledgments}

HP would like to thank Dr. Aaron Tohuvavohu for the additional guidance and recommendations throughout this project. This work made use of the Illinois Campus Cluster, a computing resource that is operated by the Illinois Campus Cluster Program (ICCP) in conjunction with the National Center for Supercomputing Applications (NCSA) and which is supported by funds from the University of Illinois at Urbana-Champaign. This work was partially supported by the Center for AstroPhysical Surveys (CAPS) at NCSA). HP and GN's work on this project is funded by AST-2206195. GN also gratefully acknowledges NSF support from NSF CAREER grant AST-2239364, supported in-part by funding from Charles Simonyi, OAC-2311355, AST-2432428, as well as AST-2421845 and funding from the Simons Foundation for the NSF-Simons SkAI Institute. GN is also supported by the DOE through the Department of Physics at the University of Illinois, Urbana-Champaign (\# 13771275), and support from the HST Guest Observer Program through HST-GO-16764.

{\it Software}: \textsc{afterglowpy} \citep{ryan_gamma-ray_2020}, \textsc{Astropy} \citep{astropy:2013, astropy:2018, astropy:2022},   
\textsc{Matplotlib} \citep{matplotlib}, \textsc{Numpy} \citep{numpy}, \textsc{Pandas} \citep{pandas1, pandas2}, \textsc{Scipy} \citep{2020SciPy-NMeth},  \textsc{Sncosmo} \citep{sncosmo}, and \textsc{label-lines} \citep{labellines}.

\bibliographystyle{mnras}
\bibliography{srcs} 

\appendix 

\counterwithin{figure}{section}

\section{Considering Other Photometric Regimes}

\subsection{Expected UVEX Photometry} \label{sec:uvex}

Given the enhancement in the bluer bands, we also considered the future NASA medium explorer UVEX, the UltraViolet EXplorer, which will observe in two ultraviolet bands \citep{Kulkarni_2021uvex}; however, this instrument is not sensitive enough to capitalize on this enhancement with its current expected $5-\sigma$ depth. 

\begin{figure}[h!]
    \centering    \includegraphics[width=\textwidth, scale=0.75]{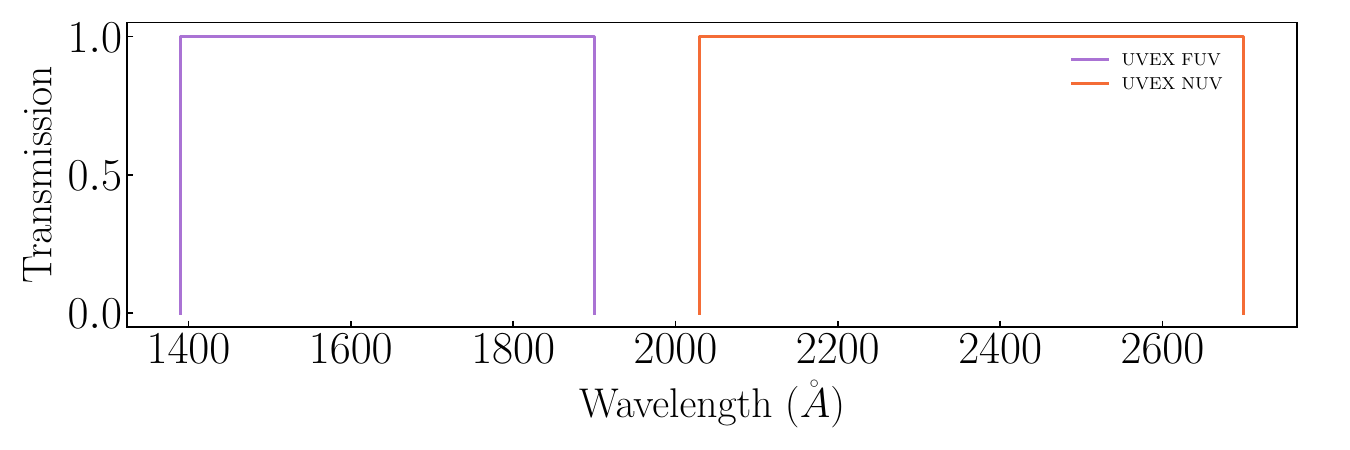}
    \caption{The tophat bandpasses used in Figure \ref{fig:uvex}.}
    \label{fig:uvex_transmission}
\end{figure}

\begin{figure}
    \centering    \includegraphics[width=\textwidth, scale=0.75]{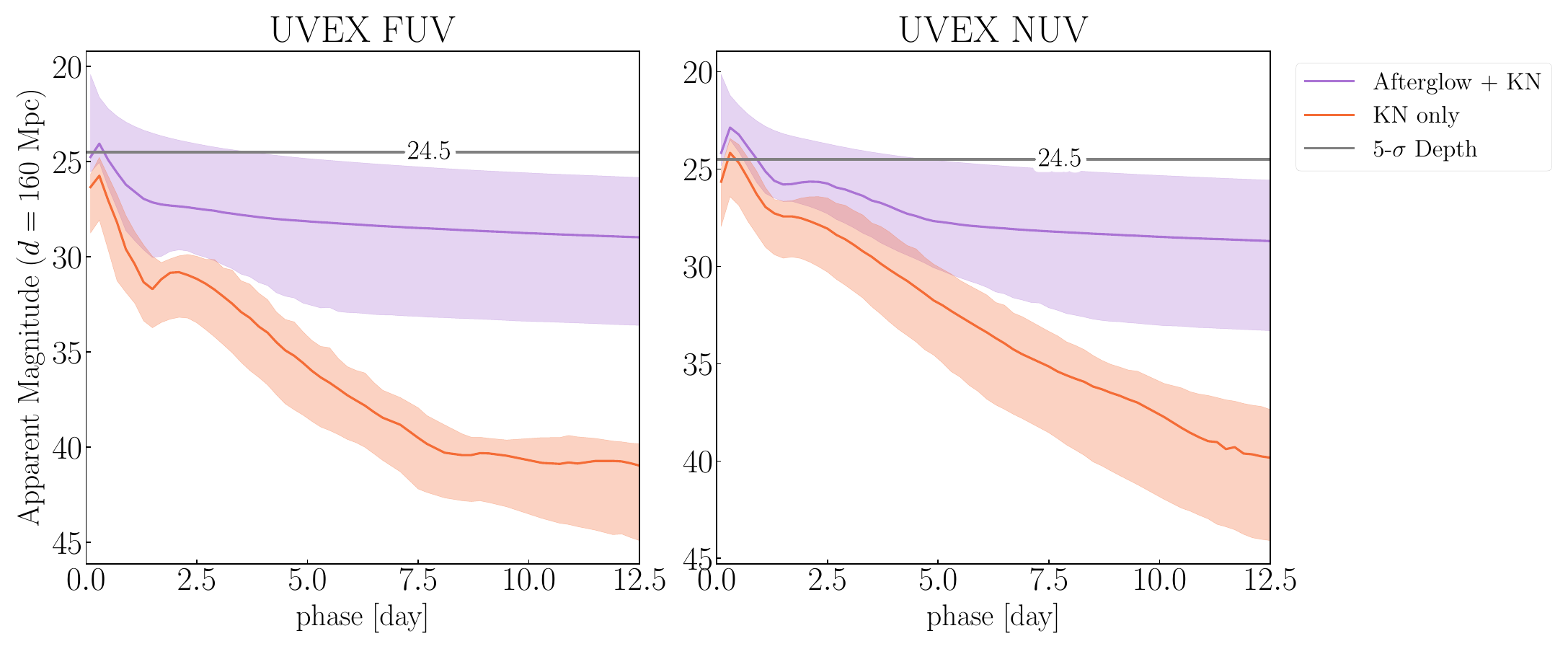}
    \caption{Like Figure \ref{fig:lsst}, but light curves in the two UVEX bands.}
    \label{fig:uvex}
\end{figure}

For this work, we assumed perfect transmission for the listed bandwidth and construct tophat bandpasses for UVEX's FUV and NUV filters \citep{Kulkarni_2021uvex}, as shown in \ref{fig:uvex_transmission}.  FUV and NUV span $1390 – 1900 \AA$ and $2030–2700 \AA$, respectively, with sensitivities of 24.5 in both bands.  As in Figure \ref{fig:lsst}, we obtain the median along with the 16th and 84th percentile light curves from the simulated sample in both UVEX filters, as shown in Figure \ref{fig:uvex}.  Despite the large enhancement relative to the KN, the anticipated sensitivity of a single exposure of UVEX will only be sufficient for very nearby events.  Additionally, this enhancement is so significant that much of the KN evolution is obstructed. While more accurate bandpasses have been reported for UVEX\footnote{\url{https://www.uvex.caltech.edu/page/for-astronomers}}, this does not change our conclusions, as reducing the perfect transmission to be more representative of the true bandpasses further exacerbates the difficulty of observing these faint events.

The interpolated KN SEDs in these ultraviolet bands experienced unphysical peaks due to noise in the grids used to build the interpolator. The grids were produced using the \texttt{POSSIS} code which can experience large Monte Carlo noise at epochs of small flux.  For the events in which an unphysical peak occurred, they were isolated to singular points in the phase-wavelength plane with $t > 10 \ \rm day$, and the sharp features in the SEDs created gaps in the synthetics light curves. These gaps were then filled with linearly interpolated values, as the overall trend of the light curve at phases around the gaps was smooth. Additionally, relative to the peak of the KN light curve the emission has decreased by 10-15 magnitudes, or by a factor of $10^{4-6}$ in flux and is too faint to be detected by a telescope. The resulting light curves can be seen in Figure \ref{fig:uvex}. 

\subsection{Infrared Observations with {\it JWST} and {\it RST}}

While the previous discussion was focused on leveraging the AG emission to enhance the UV/optical brightness of a NSM, this enhancement can obscure the temporal evolution of the KN in bluer bands. As one moves to bluer bands, from $g$-band in \ref{fig:lsst} to FUV \ref{fig:uvex}, the afterglow contribution becomes more dominant. We now consider the infrared. 

\begin{figure*}
    \centering    \includegraphics[width=\textwidth, scale=0.75]{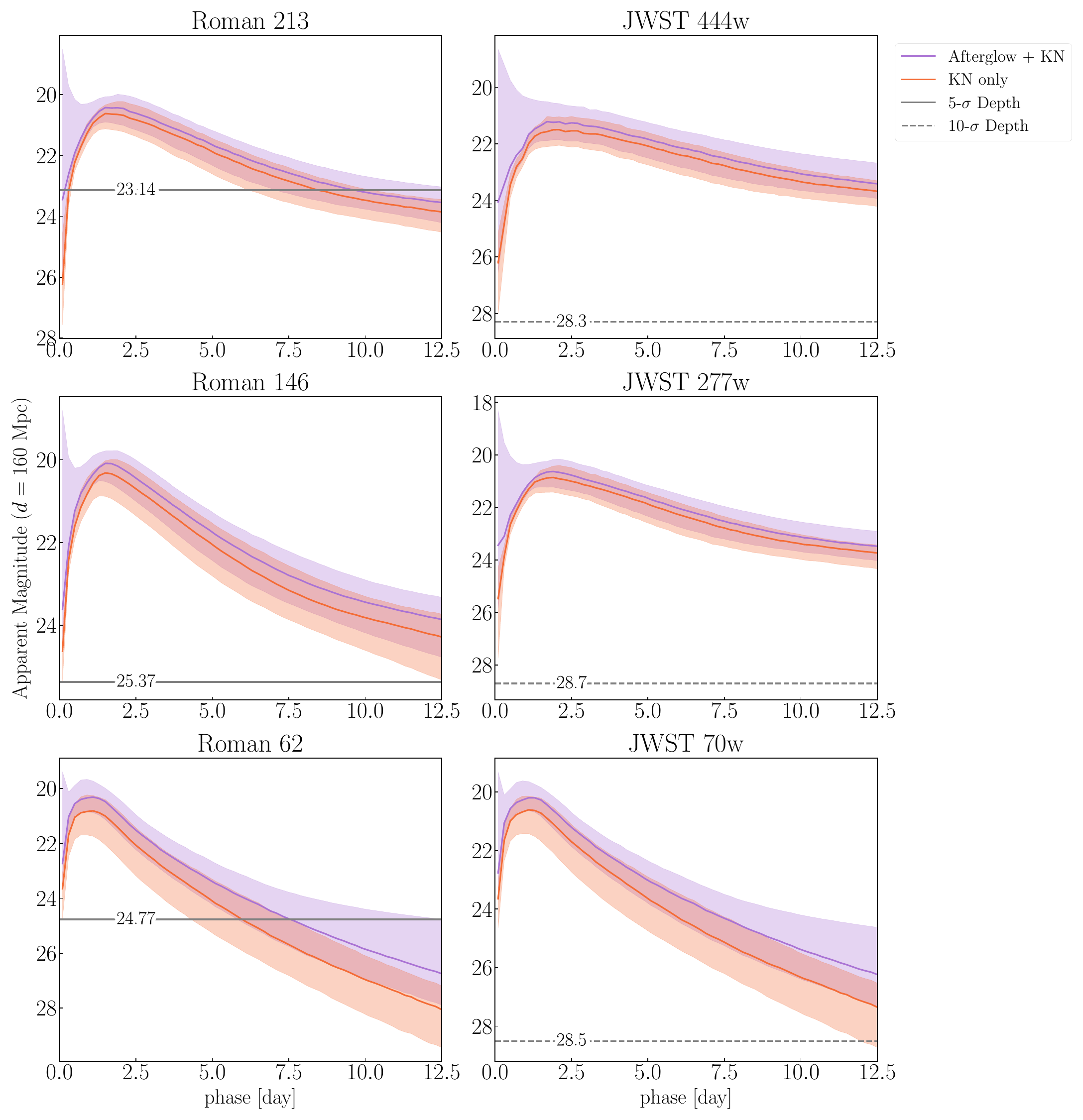}
    \caption{Like Figure \ref{fig:lsst}, but light curves in three Roman and three JWST passbands.}
    \label{fig:jwst}
\end{figure*}

Here, we consider the current {\it James Webb Space Telescope}'s ({\it JWST}) Near-Infrared Camera (NIRCam) and upcoming {\it Roman Space Telescope} ({\it RST}) Wide Field Instrument (WFI) for their near-infrared (NIR) imaging capabilities.  We perform the same calculations as in Section \ref{sec:simulate} but in 3 NIRCam bands and 3 WFI bands. We select F070W and F444W to include the extremes in observed wavelength for NIRCam, with F277W being an intermediate filter. Additionally, observations in F444W were used to identify GRB230307A's red counterpart as a kilonova \citep{Levan_2023_grb23}.  The {\it RST} F062, F146, and F213 bands were shown to be able to detect a range of kilonova models \citep{andreoni_enabling_2023}. For the limiting magnitudes shown in Figure \ref{fig:jwst}, we use the 10-$\sigma$ point source depths as listed in the JWST User Documentation\footnote{As of August 1st, 2024: \url{https://jwst-docs.stsci.edu/jwst-near-infrared-camera/nircam-performance/nircam-sensitivity}}, which accounts for an exposure of $10 \ \rm ks$, and 5-$\sigma$ point source limiting magnitude as listed in the Roman Space Telescope Wide-Field Instrument and Observatory Performance\footnote{As of June 4th, 2024: \url{https://roman.gsfc.nasa.gov/science/WFI_technical.html}}.

As suggested by \citet{Bartos2016_jwstKN, andreoni_enabling_2023, Rose_2025_romanElasticc}, {\it JWST} and {\it RST} are fully capable of observing KNe, thus making them great follow-up instruments. 
 {\it RST} will have a public survey component, the High Latitude Time-Domain Core Community Survey, which will be a nice compliment to LSST. 
 In these bands, we see that the additional emission from the AG is not needed to boost the event over the sensitivity limits for events at a distance of $160$ Mpc and the dominant emission is from the KN. Additionally, at the times for which the AG is dominant in the bluer band ($t < 10$ d), the KN emission dominates in the IR, thus allowing for follow up in these bands with minimal contamination from AG emission.  However, in extreme cases (0-16th percentile of events, which are those brighter than the upper bounds in Figure \ref{fig:jwst}), where very energetic events ($E_{0} \gtrsim 10^{52}$ erg) viewed very near on-axis, the AG can outshine the kilonova even in the IR.   

\end{document}